\documentclass[12pt,usenames,dvipsnames,a4paper]{article}

\usepackage{soul}

\usepackage[utf8]{inputenc}
\usepackage[T1]{fontenc}

\usepackage{changepage}
\usepackage{amsmath,amsfonts,amssymb}
\usepackage{epsf,bbold,stmaryrd}
\usepackage{mathrsfs}
\usepackage{appendix}
\usepackage{caption}
\usepackage{color}
\usepackage{datetime}
\usepackage{float}
\usepackage{graphicx}
\usepackage[colorlinks,hyperindex,breaklinks]{hyperref}
\hypersetup{pageanchor=false,citecolor=red,urlcolor=red}
\usepackage{indentfirst}
\usepackage[numbers,square,comma,sort&compress,merge]{natbib} 
\usepackage{subcaption}
\usepackage{mathtools}
\usepackage{ytableau}
\usepackage{tabu}
\usepackage{esvect}

\usepackage{calc}
\usepackage{bm}

\allowdisplaybreaks

\def\a{\alpha}

\def\c{\chi}

\def\d{\delta}
\def\D{\Delta}

\def\f{\frac}

\def\G{\Gamma}

\def\l{\left}

\def\mc{\mathcal}

\def\m{\mu}
\def\n{\nu}

\def\p{\partial}

\def\r{\right}
\def\s{\sigma}

\def\z{\zeta}

\def\be{\begin{equation}}
\def\ee{\end{equation}}

\def\bea{\begin{eqnarray}}
\def\eea{\end{eqnarray}}

\def\ba{\begin{array}}
\def\ea{\end{array}}

\def\bc{\begin{center}}
\def\ec{\end{center}}

\def\bl{\begin{flushleft}}
\def\el{\end{flushleft}}

\def\br{\begin{flushright}}
\def\er{\end{flushright}}

\def\bi{\begin{itemize}}
\def\ei{\end{itemize}}

\def\bt{\begin{tabular}}
\def\et{\end{tabular}}

\makeatletter

\newsavebox\myboxA
\newsavebox\myboxB
\newlength\mylenA

\newcommand*\xoverline[2][0.75]{%
    \sbox{\myboxA}{$\m@th#2$}%
    \setbox\myboxB\null
    \ht\myboxB=\ht\myboxA%
    \dp\myboxB=\dp\myboxA%
    \wd\myboxB=#1\wd\myboxA
    \sbox\myboxB{$\m@th\overline{\copy\myboxB}$}
    \setlength\mylenA{\the\wd\myboxA}
    \addtolength\mylenA{-\the\wd\myboxB}%
    \ifdim\wd\myboxB<\wd\myboxA%
       \rlap{\hskip 0.5\mylenA\usebox\myboxB}{\usebox\myboxA}%
    \else
        \hskip -0.5\mylenA\rlap{\usebox\myboxA}
         {\hskip 0.5\mylenA\usebox\myboxB}%
    \fi}
\makeatother

\usepackage{xspace}
\newcommand*{\ie}{i.e., }

\usepackage{xifthen}
\newcommand*\diff{\mathrm{d}} 
\newcommand*\ldiff[2][]{ \ifthenelse{\isempty{#1}}{ \diff
#2}{\diff^#1#2} \,} 
\let\limitint\int 
\renewcommand{\int}{\limitint \!} 

\begin{document}

\begin{titlepage}

\vspace{7cm}

\begin{center}
    \Large\textbf{A scaling non-compact QCD axion}
\end{center}

\vspace{1cm}

\begin{center}
\textsc{Georgios K. Karananas,$^{\star,\dagger}$ Mikhail
Shaposhnikov$^\ddagger$}
\end{center}

\begin{center}
\it {$^\star$Max-Planck-Institut f\"ur Physik\\
Boltzmannstra{\ss}e 8, 85748, Garching bei M\"unchen, Germany\\
\vspace{.4cm}
$^\dagger$Arnold Sommerfeld Center\\
Ludwig-Maximilians-Universit\"at M\"unchen\\
Theresienstra{\ss}e 37, 80333 M\"unchen, Germany\\
\vspace{.4cm}
$^\ddagger$Institute of Physics \\
\'Ecole Polytechnique F\'ed\'erale de Lausanne (EPFL) \\ 
CH-1015 Lausanne, Switzerland\\
\vspace{.4cm}
}
\end{center}

\begin{center}
\small
\texttt{\small georgios.karananas@physik.uni-muenchen.de}  \\
\texttt{\small mikhail.shaposhnikov@epfl.ch} 
\end{center}

\vspace{2cm}

\begin{abstract}

We present a dynamical mechanism for the erasure of inflationary isocurvature
perturbations of the non-compact QCD axion. The key ingredient is an
early-time runaway exponential potential, which drives the axion onto the
well-known scaling cosmological attractor after inflation. Once on the
attractor, the axion tracks the dominant component of the Universe, radiation,
and isocurvature modes are erased even if the field is effectively massless
during inflation. When the QCD potential turns on, the axion carries nonzero
velocity, and kinetic misalignment can become operative. The exponential
potential induces residual CP violation, potentially accessible to future
electric dipole moment searches. This mechanism requires that the axion be
effectively non-compact over the field range relevant for its
post-inflationary evolution.

\end{abstract}

\end{titlepage}

\section{Introduction}
 
The most elegant solution to the strong CP puzzle promotes the physical CP
violating parameter $\theta_{\rm phys}$ to a dynamical
field~\cite{Peccei:1977hh,Weinberg:1977ma,Wilczek:1977pj}---the axion---which
relaxes toward a CP-conserving minimum, while at the same time providing an
excellent Dark Matter (DM) candidate. This economy is one of the main reasons
why the axion idea has remained compelling for so long.

Standard implementations\,\footnote{Other important approaches rely on string
theory~\cite{Witten:1984dg}, extra dimensions~\cite{Reece:2024wrn}, or gauge
symmetry~\cite{Dvali:2005an,Dvali:2013cpa,Dvali:2017mpy,Dvali:2022fdv}.} of
this idea rely on a spontaneously broken Peccei-Quinn (PQ) symmetry. This
logic is so deeply ingrained that it is easy to lose sight of what is actually
necessary in order to solve the strong CP problem. In fact, the only thing
required is a sufficiently light dynamical degree of freedom coupled to the
QCD topological density. 

Once this is appreciated, the natural question is whether the strong CP puzzle
can be solved outside the PQ framework. If the answer is in the affirmative,
then one may be led to axion cosmologies that differ from the conventional
paradigm; the well-known cosmological
problems~(see~\cite{Sikivie:2006ni,Marsh:2015xka,DiLuzio:2020wdo,OHare:2024nmr}
for reviews)~are closely tied to the axion's PQ origin: a potential domain
wall problem~\cite{Sikivie:1982qv} when the symmetry breaks after inflation,
or stringent isocurvature constraints from Cosmic Microwave Background (CMB)
radiation observations~\cite{Planck:2018jri} when it breaks before or during
inflation. 

Among these issues, axion isocurvature perturbations constitute an obstacle to
a cosmological phenomenology consistent with observations. In the
pre-inflationary PQ-breaking scenario, the axion---which in its vanilla
realizations, and assuming a standard cosmological evolution, is effectively
massless during inflation---acquires fluctuations that source isocurvature
modes. These are tightly constrained by measurements of the CMB and, in the
usual picture, require tuning of the initial axion value (equivalently, the
misalignment angle). 

A~\emph{statistical} solution to the isocurvature problem, inspired
by~\cite{Linde:1990yj}, was proposed in~\cite{Karananas:2025uhy}. It relies on
abandoning the usual PQ origin of the axion and instead considering a
\emph{non-compact} field, \ie one not restricted to a $2\pi$ interval. In that
case, inflation stretches the axion fluctuations to superhorizon scales, so
that when the QCD potential turns on the field samples many distinct QCD vacua
across the observable Universe. The axion density on CMB scales is averaged
over many uncorrelated branches, and the would-be large-scale isocurvature
modes are ``washed-out.'' 

In this setup the axion can account for all or part of the observed DM
abundance, and it leads to a distinctive cosmology. Since no global U(1)$_{\rm
PQ}$ symmetry is invoked, domain walls form but no cosmic strings are present.
A small non-QCD contribution to the axion potential---taken
in~\cite{Karananas:2025uhy} to be a mass term---is
required~\cite{Sikivie:1982qv} to lift the degeneracy between neighboring
vacua and ensure wall annihilation before Big Bang Nucleosynthesis, thereby
avoiding cosmological catastrophe. This in turn implies both nonvanishing
strong-CP violation and a stochastic gravitational-wave background at
nanohertz frequencies, offering correlated targets for electric dipole moment
searches~\cite{Anastassopoulos:2015ura,Grasdijk:2020ihi} and pulsar timing
arrays~\cite{Babak:2024yhu}.

The present work follows the same general spirit, in that the axion is taken
to be non-compact with a non-QCD piece in its potential. The crucial
difference is that here this extra contribution is not a mass term, but an
exponentially decaying potential---well known from studies of
quintessence---which supports a scaling attractor
solution~\cite{Wetterich:1987fm, Ratra:1987rm}~(for detailed studies
see~\cite{Ferreira:1997hj,Copeland:1997et}). In this way, we arrive at
a~\emph{dynamical} solution to the isocurvature problem. These perturbations
are ``erased'' through the time evolution of the homogeneous mode itself
rather than ``washed-out'' through averaging over many patches. The case of an
exponential, but not steep potential that does not support an attractor has
been already considered in~\cite{Karananas:2026pov}.

For the erasure to be achieved, the attractor must be reached well before the
QCD epoch. As we shall show, this can indeed happen under very mild
assumptions on the onset of the attractor dynamics. Once the field has joined
the scaling solution, memory of initial conditions is lost, the subsequent
evolution becomes universal and fixed, and the energy density of the field
tracks that of the dominant background. In other words, the axion is no longer
an independent species, but is enslaved to the background expansion and
behaves as a fixed fraction of the dominant constituent; in the era relevant
for us, that constituent is radiation. As a result, the attractor mechanism
provides a genuine dynamical resolution of the isocurvature problem.

For an axion on the scaling attractor, its motion is no longer arbitrary, as
the field rolls with a velocity fixed by the background expansion. In
particular, unlike in the conventional misalignment picture, the axion is not
frozen until the QCD epoch. Instead, it is already rolling when the QCD
potential emerges, carrying a definite and nonvanishing kinetic energy
inherited from the attractor evolution. The onset of QCD effects therefore
does not set the axion in motion from rest. Rather, it interrupts an already
established scaling trajectory and forces the field into a new dynamical
regime in which the periodic QCD potential competes with the pre-existing
motion. This is precisely the origin of the kinetic misalignment
mechanism~\cite{Co:2019jts}, and the subsequent abundance is controlled by the
velocity with which the field enters the QCD potential.

It is important to note that to not overproduce DM when kinetic misalignment
is operative, the axion must be captured by the QCD potential no later than
temperatures of order 1~{\rm GeV}, see~\cite{Barman:2021rdr}. This already
implies that the axion velocity at QCD onset cannot be arbitrarily large. In
the present setup, that velocity is controlled by the parametric size of the
exponential slope, which is a free parameter of the theory. Moreover, the very
presence of a non-QCD contribution to the axion potential generically induces
residual CP violation in the strong sector, since the total potential is
shifted away from the parity-preserving point.

Note that unlike the statistical mechanism of~\cite{Karananas:2025uhy}, the
present setup does not predict a cosmological domain wall network. The reason
is that the attractor homogenizes the axion over all superhorizon patches
before the QCD potential becomes relevant, so that the observable Universe is
captured into a single branch. Thus, no gravitational-wave signal associated
with wall collapse is present here. This is one of the sharpest
phenomenological distinctions between this mechanism and the earlier
statistical one.

A further important issue concerns the stability of the vacuum in which the
axion eventually settles. Because the potential has the form of a tilted
washboard (see Fig.~\ref{fig:Potential}), there always exists a lower-lying
vacuum to which the field may, in principle, tunnel. It is therefore natural
to worry about the fate of the state selected cosmologically. As we shall see,
however, the corresponding tunneling rate is negligibly small, since the
energy splitting between neighboring vacua is tiny.

This paper is organized as follows. In Sec.~\ref{sec:attractor}, we briefly
review the basic properties of the scaling attractor. In Sec.~\ref{sec:QCD},
we analyze the dynamics of the QCD axion in the presence of the attractor
potential. In Sec.~\ref{sec:DM_abundance}, we discuss the dark matter
abundance, and in Sec.~\ref{sec:CP_violation}, the residual CP violation. In
Sec.~\ref{sec:stability}, we examine the stability of the selected vacuum. In
Sec.~\ref{sec:EFT}, we assess the validity of the effective field theory. In
Sec.~\ref{sec:exp_origins}, a possible gravitational origin of the exponential
potential is presented. We conclude in Sec.~\ref{sec:conclusion}.

\section{Attractor basics}
\label{sec:attractor}

\subsection{The scaling solution}

We begin with the Lagrangian for the axion field $a$ that reads
\be
\label{eq:lagrangian_attractor}
\mc L = - \f 1 2 (\p_\m a)^2 - V_{\rm exp} +
\f{a}{f_a}\f{g_s^2}{32\pi^2}G^b_{\m\n}\widetilde G^{b\,\m\n} \ , 
\ee
where
\be
\label{eq:exp_potential}
V_{\rm exp} = \Lambda^4 e^{-a/F} \ ,
\ee
and $\Lambda, F, f_a$ are free dimensionful parameters; as usual, $G_{\m\n}$
and $\widetilde G_{\m\n}$ are the QCD field strength and its dual, and $g_s$
is the strong coupling. At early times, the QCD piece is subdominant wrt
$V_{\rm exp}$, so can be safely neglected for the discussions of this and the
next subsection. A possible origin of the exponential
potential~(\ref{eq:exp_potential}) was discussed in~\cite{Karananas:2026pov},
see also Sec.~\ref{sec:exp_origins}.

The cosmological dynamics of a scalar field with exponential potential is well
known~\cite{Wetterich:1987fm, Ratra:1987rm, Ferreira:1997hj,Copeland:1997et}.
For sufficiently steep exponentials there exists a scaling attractor on which
the scalar tracks the dominant component of the Universe: its energy density
scales in the same way as the background and stays a fixed fraction of the
total. As we are interested in the post-inflationary epoch preceding QCD, we
focus from the outset to radiation domination, for which the scale factor is
$R\propto t^{1/2}$ and the Hubble parameter $H = 1/2t$.

On the attractor the field grows logarithmically, $\bar a = 2F\log(t/t_0)+{\rm
const.}$, with $t_0$ an arbitrary reference time; therefore, its velocity is
\be
\label{eq:bar_a_attractor}
\dot{\bar a} = 4FH \ .
\ee
Inserting this into the equation of motion for $a$ locks the exponential to
the expansion rate of the Universe
\be
\label{eq:Lambda_attractor}
\Lambda^4 e^{-\bar a/F} = 4F^2 H^2 \ ,
\ee
a relation used repeatedly below. The scalar then tracks the radiation
background, with abundance 
\be
\label{eq:Omega_attractor}
\Omega_a = \f{\rho_a}{3M_{\rm Pl}^2 H^2} = 4\l(\f{F}{M_{\rm Pl}}\r)^2 \ . 
\ee
The attractor solution exists for $\Omega_a<1$, \ie $F<M_{\rm Pl}/2$.

\subsection{Erasure of isocurvature perturbations}

For $F\ll M_{\rm Pl}$ the field carries a negligible fraction of the energy
density, $\Omega_a\ll1$, and evolves on a fixed radiation background. A
perturbation of the axion is then an isocurvature one, since the metric and
radiation fluctuations are unaffected by the axion
sector~\cite{Ferreira:1997hj}. If the axion is an effectively massless
spectator during inflation, it acquires superhorizon fluctuations of amplitude
\be
\label{eq:delta_inflation}
\d a_I \simeq \f{H_I}{2\pi} \ ,
\ee
with $H_I$ the Hubble parameter during inflation. As this can be much larger
than $F$, different patches can in principle be far from the neighbourhood of
the attractor. This, however, does not invalidate the mechanism. For an
exponential potential the scaling solution is a global late-time
attractor~\cite{Ferreira:1997hj,Copeland:1997et}. Every patch is thus driven
to the same fixed point, and what survives is the universal adiabatic tracking
value. 

Let us denote by $t_1$ (temperature $T_1$) the time at which the field has
joined the attractor, and by $t_2$ (temperature $T_2$) the later time at which
QCD effects become important and the scaling solution breaks down, see
Sec.~\ref{sec:QCD}. Decomposing the perturbations of the field into Fourier
modes $\d a_k$, the superhorizon ones decay as~\cite{Ferreira:1997hj}
\be
\label{eq:delta_a_decay}
\d a_k \propto \sqrt{\f{T_2}{T_1}} \ .
\ee
That the scaling attractor erases isocurvature in this way was first pointed
out in~\cite{Ferreira:1997hj}; see~\cite{Abramo:2001mv} for a gauge-invariant
treatment of the decay.

The isocurvature mode is therefore efficiently erased provided
\be
\label{eq:iso_condition}
\f{H_I}{2\pi F}\,\sqrt{\f{T_2}{T_1}}\ll 10^{-5} \ ,
\ee
with the right-hand side being the current CMB isocurvature
bound~\cite{Planck:2018jri}. For $H_I/(2\pi F)\sim\mc O(10)$ this requires
only $T_1/T_2\gtrsim 10^{12}$, comfortably below the $\sim\mc
O(10^{14}-10^{15})$ available between (high-scale) reheating and the QCD
epoch. The axion thus reaches the QCD onset in the universal adiabatic
tracking configuration, with the primordial isocurvature component heavily
suppressed and no memory of initial conditions.

\section{Onset of QCD and breakdown of the attractor}
\label{sec:QCD}

At early times the evolution is well approximated by the scaling solution
discussed in the previous section. At later times, however, QCD effects induce
the usual potential for the axion, which we shall approximate as
\be
V_{\rm QCD} = \c(T)\l(1 - \cos\l(\f{a}{f_a} - \bar \theta \r)\r) \ ,
\ee
where $\chi(T)$ is the topological susceptibility of the vacuum, $f_a$ the
axion decay constant and $\bar \theta$ the theta parameter. Therefore, the
total Lagrangian reads
\be
\mc L = - \f 1 2 (\p_\m a)^2 - V_{\rm exp} - V_{\rm QCD} \ ,
\ee
with $V_{\rm exp}$ given in~(\ref{eq:exp_potential}). 

The attractor ceases to be a good description once the QCD force becomes
comparable to the exponential slope, \ie 
\be
|V'_{\rm exp}| \sim|V'_{\rm QCD}| \ .
\ee
The onset of QCD effects is the time $t_2$ at which this condition is first
saturated
\be
\label{eq:break_condition_general}
\f{\Lambda^4}{F}e^{-a_2/F} \sim \f{\c(T_2)}{f_a} \ ,
\ee
where $a_2 = a(t_2)$.

Using the attractor relation~(\ref{eq:Lambda_attractor}) during radiation
domination, the breakdown condition becomes
\be
\label{eq:break_condition}
\f{m_a(T_2)}{H_2} \sim 2\sqrt{\f{F}{f_a}} \ ,
\ee
where
\be
\label{eq:axion_mass}
m_a^2(T) =  \f{\c(T)}{f_a^2} \ .
\ee

At the same time
\be
\label{eq:Hubble_RD}
H(T)=\f{\pi}{3}\sqrt{\f{g_\ast(T)}{10}}\f{T^2}{M_{\rm Pl}} \ ,
\ee
while the temperature-dependent axion mass may be approximated
as~\cite{Marsh:2015xka,OHare:2024nmr}\,\footnote{We took the standard
high-temperature behavior $\c(T)\propto T^{-8}$, \ie $m_a(T)\propto T^{-4}$;
different exponents inferred from lattice studies would only affect the result
through small fractional powers, so the sensitivity is mild. }
\be
\label{eq:axion_mass_T}
m_a(T)\simeq
\begin{cases}
m_a\l(\f{T_{\rm QCD}}{T}\r)^4 & {\rm for}~T > T_{\rm QCD} \ ,\\
m_a & {\rm for}~T < T_{\rm QCD} \ ,
\end{cases}
\ee
with
\be
\label{eq:axion_mass_0}
m_a = 5.7\times 10^{-12}\,{\rm GeV}
\l(\f{10^9~{\rm GeV}}{f_a}\r) \ .
\ee

Assuming $T_2>T_{\rm QCD}$, one then finds
\be
\label{eq:T2_estimate}
T_2 \simeq
\l[\f{m_a^2 f_a T_{\rm QCD}^8 M_{\rm Pl}^2} {g_\ast(T_2)F}\r]^{\f{1}{12}} \ ,
\ee
up to factors of order one.

Using $T_{\rm QCD}\simeq 150~{\rm MeV}$, $M_{\rm Pl}\simeq 2.4\times
10^{18}~{\rm GeV}$, and $g_\ast(T_2)\simeq 60$, one finds for the breakdown
temperature
\be
T_2 \sim 2~{\rm GeV}
\l(\f{4\times10^{11}~{\rm GeV}}{F}\r)^{\f 1{12}}
\l(\f{10^9~{\rm GeV}}{f_a}\r)^{\f 1{12}} \ .
\ee

\begin{figure}[!t]
    \centering
    \includegraphics[scale=.22]{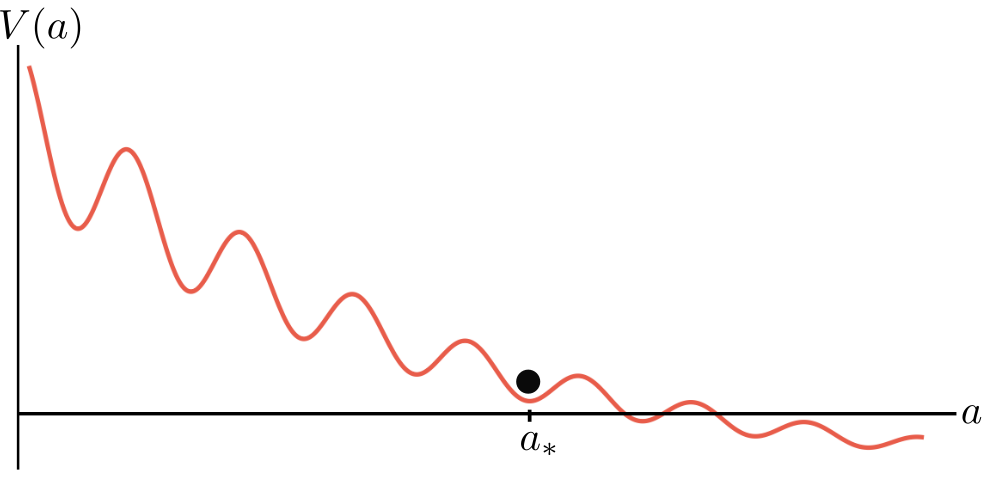}
    \caption{Tilted washboard potential for the non-compact axion in the
    over-barrier regime. $a_\ast$ denotes the field value in the branch in
    which the axion gets eventually captured. For $F>f_a$, the kinetic energy
    inherited from the attractor is large enough for the axion to cross
    several QCD barriers before being captured. In the under-barrier regime
    where $F<f_a$, the field is trapped in the first available well. After the
    cosmological constant of the selected vacuum is adjusted to its observed
    value, lower-lying branches have negative energy.}
    \label{fig:Potential}
\end{figure}

It is important to stress that the departure from the attractor does not imply
the immediate trapping of the field. At $t_2$, the full potential merely
starts to develop minima. Whether the field is immediately captured or instead
overshoots the newly-formed minima depends on its kinetic energy relative to
the height of the QCD barriers. Indeed,
using~(\ref{eq:bar_a_attractor},\ref{eq:break_condition},\ref{eq:axion_mass}),
we find
\be
\label{eq:KE_over_barrier}
\f{\dot a^2}{\c(T_2)} \sim \,\f{F}{f_a} \ ,
\ee
so the dynamics splits into two regimes according to the ratio $F/f_a$:

\begin{itemize}

\item[\emph{i)}] ``Over-barrier,''~$F>f_a$:~The kinetic energy exceeds the
barrier, the field overshoots the QCD minima and traverses several branches
before capture, see Fig.~\ref{fig:Potential}. The dynamics is of the
kinetic-misalignment type and the abundance is controlled by $F$. 

\item[\emph{ii)}] ``Under-barrier,''~$F<f_a$:~The kinetic energy is below the
barrier and the field is captured within a single well essentially at $t_2$.
Once the exponential is negligible it behaves as an ordinary misalignment
field. It remains frozen until $m_a(T)\sim 3H$, and then starts oscillating
from a generic $\mc O(1)$ initial angle. The capture temperature and abundance
are those of standard misalignment set by the decay constant $f_a$, and are
independent of $F$. Consistently, the estimate~(\ref{eq:Tstar_estimate}) gives
$T_\ast>T_2$ in this regime, signalling that kinetic misalignment does not
apply. 

\end{itemize}

It is also worth emphasizing that the mechanism requires a genuinely
non-compact field range. From the attractor
velocity~(\ref{eq:bar_a_attractor}), the total field excursion between the
onset of the attractor and the breakdown time is
\be
\D a_{\rm tot} \sim 4F \log\f{T_1}{T_2} \ .
\ee
For a high reheating scale, this can easily amount to $\mc O (100)F$, so that
the field traverses
\be
\label{eq:axion_excursion}
\frac{\D a_{\rm tot}}{2\pi f_a} \sim \frac{2F}{\pi f_a}\log\f{T_1}{T_2} \ ,
\ee
QCD periods before the periodic potential becomes important. This makes
explicit why the axion must be non-compact (at least) over the field range
relevant for its post-inflationary evolution.

Despite the large excursion~(\ref{eq:axion_excursion}), the mechanism does not
produce domain walls. That would require different Hubble patches to be
captured into different QCD branches, which in turn requires an axion
variation at least of order $2\pi f_a$ across the observable Universe at the
time the QCD potential becomes relevant. However, after the attractor has
damped the isocurvature component, the remaining fluctuation is only the
adiabatic one $\d a_{\rm ad}=\dot{\bar a}\d t$.
Using~(\ref{eq:bar_a_attractor}) and $H\d t\sim\z$, with $\z\approx 10^{-5}$
the curvature perturbation, we conclude that 
\be
\d a_{\rm ad} \sim F \z \ll 2\pi f_a \ ,
\ee
for the parameter range of interest; therefore, all observable patches are
captured into the same QCD branch.

\section{Dark Matter abundance}
\label{sec:DM_abundance}

\subsection{Over-barrier regime, $F>f_a$.} 

Kinetic misalignment is operative once the field keeps rolling even after the
full potential has developed minima. The final axion abundance is then
determined by the trapping time $t_\ast \gtrsim t_2$, rather than by the
breakdown time $t_2$. Following the standard estimate~\cite{Co:2019jts},
trapping occurs when
\be
\dot a(T_\ast) \sim 2m_a(T_\ast)f_a \ .
\ee

From~(\ref{eq:bar_a_attractor}), we find that for radiation domination $\dot a
\sim 4F H$, therefore
\be
\label{eq:trap_condition_2}
\f{m_a(T_\ast)}{H_\ast} \sim 2\f{F}{f_a} \ .
\ee

Using~(\ref{eq:Hubble_RD}) and~(\ref{eq:axion_mass_T}), for $T_\ast >
T_{\rm QCD}$ this gives
\be
\label{eq:Tstar_estimate}
T_\ast \simeq 
\l[\f{m_a^2 f_a^2 T_{\rm QCD}^8 M_{\rm Pl}^2} 
{g_\ast(T_\ast)F^2}\r]^{\f{1}{12}} \ .
\ee
Notice that due to~(\ref{eq:axion_mass_0}), $T_\ast$ is (practically)
independent of $f_a$.

As we did before, we take $T_{\rm QCD}\simeq 150~{\rm MeV}$, $M_{\rm Pl}\simeq
2.4\times 10^{18}~{\rm GeV}$, and $g_\ast(T_\ast)\simeq 60$, to get an
estimate for the capture temperature
\be
T_\ast \sim 1.3~{\rm GeV} 
\l(\f{4\times 10^{11}~{\rm GeV}}{F}\r)^{\f 1 6} \ .
\ee
Comparing~(\ref{eq:T2_estimate}) and~(\ref{eq:Tstar_estimate}), one obtains
\be
\label{eq:T2_over_Tstar}
T_\ast \sim \l(\f{f_a}{F}\r)^{\f{1}{12}}T_2 \ , 
\ee
meaning that for $F > f_a$, one has $T_\ast\lesssim T_2$, as required for the
field to first depart from the attractor and only later become trapped. The
hierarchy is however (very) mild, reflecting the fact that the interval
between $t_2$ and $t_\ast$ is short.

The present-day axionic abundance follows simply from entropy conservation
\be
\label{eq:rho_over_s}
\f{\rho_a}{s} \sim \f{m_a n_a(T_\ast)}{s(T_\ast)} \ ,
\ee
where the axion number density at trapping is
\be
\label{eq:nastar}
n_a(T_\ast)\sim 2 \dot a(T_\ast)f_a  \ .
\ee
A convenient estimate is\,\footnote{ Using~(\ref{eq:trap_condition_2})
together with~(\ref{eq:Hubble_RD}), one has
\be
n_a(T_\ast)\sim 2\dot a(T_\ast)f_a \sim F H_\ast f_a
\sim \frac{F f_a T_\ast^2}{M_{\rm Pl}} \ ,
\ee
and therefore
\be
\frac{n_a(T_\ast)}{s(T_\ast)}
\sim
\frac{F f_a}{M_{\rm Pl}T_\ast} \ .
\ee
Since
\be
\frac{\rho_a}{s}\sim m_a \frac{n_a(T_\ast)}{s(T_\ast)} \ ,
\ee
and $m_af_a$ is independent of $f_a$, one finds
\be
\frac{\rho_a}{s}\propto \frac{F}{T_\ast} \ .
\ee
Finally, Eq.~(\ref{eq:Tstar_estimate}) implies $T_\ast\propto F^{-1/6}$, so
\be
\frac{\rho_a}{s}\propto F^{7/6} \ ,
\ee
which explains the scaling in~(\ref{eq:Omega_numeric}).}
\be
\label{eq:Omega_numeric}
\Omega_a h^2 \sim 0.12 \l(\f{F}{4\times 10^{11}\,{\rm GeV}}\r)^{7/6} \ ,
\ee
up to factors of order one associated with the precise trapping condition, and
the temperature dependence of $m_a(T)$.

If the axion constitutes all of the dark matter, the
estimate~(\ref{eq:Omega_numeric}) points to a characteristic scale
\be
\label{eq:F_DM}
F \sim \mc O ({\rm few})\times 10^{11}\,{\rm GeV} \ .
\ee
More generally, if the axion makes up only a fraction $W =
\Omega_a/\Omega_{\rm DM}$ of the observed dark matter, then
\be
\label{eq:F_fraction}
F \sim 4\,  W^{6/7} \times 10^{11}\,{\rm GeV} \ .
\ee

\subsection{Under-barrier regime, $F<f_a$.}

The field gets trapped within one well at $t_2$ and the subsequent evolution
is that of standard misalignment. Oscillations begin at $T_{\rm osc}$ defined
by $m_a(T_{\rm osc})\sim 3H(T_{\rm osc})$, parametrically distinct
from~(\ref{eq:Tstar_estimate}), and the present abundance is given by the
standard formula~\cite{ParticleDataGroup:2024cfk}
\be
\label{eq:Omega_standard}
\Omega_a h^2 \sim 3.2\times 10^{-5}\,\theta_i^2 
\l(\f{f_a}{10^9~{\rm GeV}}\r)^{7/6} \ ,
\ee
with $\theta_i\sim\mc O(1)$ the value at which the field freezes. The
abundance depends only on $f_a$. Note that at the crossover $F\sim f_a$ the
over-barrier result $\Omega_a h^2\propto F^{7/6}$ and the under-barrier result
$\Omega_a h^2\propto f_a^{7/6}$ scale with the same power, so the two regimes
interpolate (up to factors of $\mc O (1)$) at $F\sim f_a$.

\section{Residual CP violation}
\label{sec:CP_violation}

After the field has been captured, the residual strong CP violation is
determined by the minimum of the zero-temperature potential
\be
V(a)=V_{\rm exp}+V_{\rm QCD}\Big|_{T=0} \ .
\ee
The corresponding extremization condition is
\be
\label{eq:minimum_qcd_full}
-\f{\Lambda^4}{F}e^{-a_{\rm min}/F}
+ m_a^2 f_a \sin\l(\f{a_{\rm min}}{f_a}-\bar\theta\r)=0 \ ,
\ee
where we used $\chi(0)\simeq m_a^2 f_a^2$.

Writing
\be
a_{\rm min} = f_a(2\pi n + \bar\theta + \theta_{\rm phys}) \ ,
~~~|\theta_{\rm phys}|\ll 1 \ ,
\ee
one finds that the physical $\theta$ parameter  is
\be
\label{eq:theta_phys_general}
\theta_{\rm phys} \simeq \f{\Lambda^4 e^{-a_\ast/F}}{F f_a m_a^2} \ ,
\ee
where $a_\ast$ is the field value in the branch where the axion eventually
settles. The two regimes of~(\ref{eq:KE_over_barrier}) differ in $a_\ast$, and
hence in the size of the tilt at the selected minimum.

Importantly, the branch in which the axion settles carries vacuum energy
\be
V(a_\ast) \simeq F f_a m_a^2\,\theta_{\rm phys} \ ,
\ee
as follows from~(\ref{eq:theta_phys_general}). This is obviously far from the
observed value. The cosmological evolution selects a branch dynamically, but
says nothing about why the field should stop in a minimum with the observed
cosmological constant. Among the infinitely many lower-lying vacua, none is
preferred energetically. Therefore, the vacuum energy must be tuned to the
observed value separately---the manifestation of the cosmological constant
problem in this context. Note that in the parameter range of interest this
tuning forces all the lower-lying branches to be anti de Sitter.

\subsection{Over-barrier regime, $F>f_a$.}

To estimate $a_\ast$, we write
\be
a_\ast = a_2 + \D a \ ,
\ee
where $a_2=a(t_2)$ is the field value at the breakdown of the attractor, and
\be
\D a = \limitint_{t_2}^{t_\ast} \diff t \, \dot a  
\sim 2 F \log\f{t_\ast}{t_2} \sim \f F 3 \log\f{F}{f_a} \ ,
\ee
is the subsequent excursion before capture; as $t_2$ and $t_\ast$ only differ
slightly, we have kept the field velocity in its attractor value. 

From~(\ref{eq:theta_phys_general}), after employing
relation~(\ref{eq:Lambda_attractor}) evaluated at $t_2$ and the breakdown
condition~(\ref{eq:break_condition}), one obtains
\be
\label{eq:theta_phys_final}
\theta_{\rm phys}
\sim
\f{m_a^2(T_2)}{m_a^2}\l(\f{f_a}{F}\r)^{1/3}.
\ee
Substituting the numerical estimate for $T_2$, see~(\ref{eq:T2_estimate}),
gives
\be
\theta_{\rm phys} \sim \mc O ({\rm few})\times 10^{-11} 
\l(\f{F}{4\times10^{11}~{\rm GeV}}\r)^{\f1{3}} 
\l(\f{f_a}{10^9~{\rm GeV}}\r) \ .
\ee

One should keep in mind that this estimate carries an intrinsic $\mc O (1)$
uncertainty. The reason is that between $t_2$ and $t_\ast$ the field traverses
a number of QCD periods before finally settling, and each additional branch
changes the exponential prefactor by $e^{-2\pi f_a/F}$. For the values of
interest here this effect is parametrically small per branch, but can
accumulate to an overall order-one uncertainty in the final prediction for
$\theta_{\rm phys}$.

In the region where the axion accounts for all of the dark matter one is
naturally led to a residual strong-CP violation close to the current
experimental bounds, potentially detectable in the planned electric dipole
moment searches~\cite{Anastassopoulos:2015ura,Grasdijk:2020ihi}. In other
words, once the dark-matter abundance fixes $F$, the lower astrophysical
limit~\cite{Kim:1986ax,Turner:1989vc,Raffelt:1990yz,Caputo:2024oqc,
Fiorillo:2025gnd,Candon:2025sdm} on $f_a\gtrsim 10^{8-9}~{\rm GeV}$ prevents
$\theta_{\rm phys}$ from becoming arbitrarily small. This makes the scenario
rather predictive, as the requirement of obtaining the observed dark matter
abundance and the existence of a nonzero residual strong-CP phase become
correlated.

\subsection{Under-barrier regime, $F<f_a$.}

The field is trapped at $t_2$ with no subsequent excursion, $a_\ast\simeq
a_2$. Reading the tilt directly off the breakdown
condition~(\ref{eq:break_condition_general}), $\Lambda^4 e^{-a_2/F}/F\sim
m_a^2(T_2)f_a$, so
\be
\label{eq:theta_phys_sub}
\theta_{\rm phys}\sim \f{m_a^2(T_2)}{m_a^2} \ ,
\ee
\ie the same expression as~(\ref{eq:theta_phys_final}), \emph{without} the
factor $(f_a/F)^{1/3}$. Using~(\ref{eq:T2_estimate}), the above gives 
\be 
\theta_{\rm phys}\sim \mc O({\rm few})\times10^{-12} 
\l(\f{F}{10^8~{\rm GeV}}\r)^{\f 2 3} 
\l(\f{f_a}{10^9~{\rm GeV}}\r)^{\f 2 3} \ .
\ee 
Here, $T_2$ is generally higher than the standard oscillation temperature,
defined by $m_a(T_{\rm osc})\sim 3H(T_{\rm osc})$. The field is captured into
a single branch at $t_2$, but its subsequent evolution begins only when the
QCD mass becomes comparable to the Hubble rate.

\section{Stability of the vacuum}
\label{sec:stability}

Because the potential is a tilted washboard, the vacuum in which the axion
eventually settles is only metastable, as there always exists a lower-lying
ground state to which the field may tunnel. It is therefore important to check
that the corresponding decay rate is negligible on cosmological timescales.

The energy splitting between adjacent vacua is set by the exponential tilt. As
before, we denote by $a_\ast$ the value of the axion in the (metastable)
branch that it got captured at. The difference between two neighbouring ground
states is
\be
\D V \simeq \Lambda^4 e^{-a_\ast/F}\l(1-e^{-2\pi f_a/F}\r) 
= F f_a m_a^2\,\theta_{\rm phys}\l(1-e^{-2\pi f_a/F}\r) \ ,
\ee
where the second equality follows from~(\ref{eq:theta_phys_general});
therefore, for the two regimes of~(\ref{eq:KE_over_barrier}) we find
\be
\label{eq:DeltaV_regimes}
\D V \simeq
\begin{cases}
2\pi f_a^2 m_a^2\,\theta_{\rm phys} \ , & F > f_a \ ,\\
F f_a m_a^2\,\theta_{\rm phys} \ , & F < f_a \ .
\end{cases}
\ee
In either case $\theta_{\rm phys}\ll1$ implies $\D V\ll m_a^2 f_a^2$---this is
the regime in which the thin-wall approximation applies.

The zero-temperature decay rate is then governed by the
bounce,\footnote{Strictly speaking, after the energy of the selected vacuum is
tuned to the observed value, the lower-lying branches are anti de Sitter,
meaning that the tunneling is described by the Coleman-De Luccia
bounce~\cite{Coleman:1980aw}, rather than by its flat space counterpart.
Gravitational corrections, however, are negligible as the bubble size is much
smaller than the curvature radius associated with the lower vacuum. Therefore,
the flat space bounce suffices; if anything, gravity only suppresses the decay
further.} with Euclidean action
\be
\label{eq:S4_bounce}
S_4 \simeq \f{27\pi^2}{2}\f{\sigma^4}{(\D V)^3} \ ,
\ee
where 
\be
\label{eq:sigma_bounce}
\sigma \simeq 8 m_a f_a^2 \ .
\ee
is the tension of the bounce.

Substituting~(\ref{eq:DeltaV_regimes},\ref{eq:sigma_bounce})
into~(\ref{eq:S4_bounce}) gives
\be
S_4 \simeq \f{7\times10^3}{\pi}\f{f_a^2}{m_a^2\theta_{\rm phys}^3} \ ,
\ee
and owing to~(\ref{eq:axion_mass_0}), this becomes
\be
S_4 \simeq 6.8\times10^{73} 
\l(\f{f_a}{10^9~{\rm GeV}}\r)^4 \l(\f{10^{-10}}{\theta_{\rm phys}}\r)^3 \ .
\ee
In the under-barrier regime the splitting in~(\ref{eq:DeltaV_regimes}) is
smaller by a factor $F/(2\pi f_a)< 1$. In turn, $S_4$ is larger and the vacuum
even more stable. 

In any event, the decay rate per unit volume $V_4$ therefore behaves as
\be
\f{\G}{V_4}\sim m_a^4 e^{-S_4} \ ,
\ee
and the lifetime of the vacuum is incomparably longer than the age of the
Universe. Hence the state selected by the cosmological evolution is for all
practical purposes stable.\footnote{Note, however, that it has been argued
that consistency may require Minkowski vacua to be exactly
stable~\cite{Dvali:2011wk}.}

\section{Validity of the effective field theory}
\label{sec:EFT}

As we have seen, the regime of most interest for the present mechanism is
$F<H_I$. At first sight, this might seem to cast doubt on the effective
description, since the fluctuations of a light spectator during inflation are
of the order of the inflationary Hubble scale, see~(\ref{eq:delta_inflation}),
and need not be small compared to $F$. This conclusion is however too naive.

The relevant question is not whether $\delta a_I/F$ is small, but whether
$V_{\rm exp}$ remains negligible over the field interval explored by the
inflationary fluctuations. Let $\bar a_I$ denote the background value of the
field during inflation; then the largest value of $V_{\rm exp}$ within a
Hubble patch is for $a=\bar a_I-\delta a_I$. A sufficient condition for the
field to remain a spectator is therefore
\be
\bar \Lambda_I e^{\delta a_I/F} \ll 3M_{\rm Pl}^2H_I^2 \ ,
\ee
and combined with the requirement that it be effectively massless, $V''_{\rm
exp}<H_I^2$,
\be
\frac{\bar \Lambda_I}{F^2}e^{\delta a_I/F} < H_I^2 \ ,
\ee
with $\bar \Lambda_I = \Lambda^4 e^{-\bar a_I/F}$, shows that the validity of
the effective description is controlled by the background-dependent quantity
$\bar \Lambda_I^{1/4}$ rather than by the ratio $F/H_I$ alone, or by
$H_I/\Lambda$. Indeed, unlike a potential whose height is fixed by its
prefactor, the parameter $\Lambda$ in the exponential has no invariant meaning
by itself, as a constant shift of $a$ can be absorbed into a rescaling of
$\Lambda$. In particular, $F<H_I$ is fully under control, provided of course
that the inflationary background lies sufficiently far from the runaway
direction.

After inflation, once the field has joined the scaling solution, the size of
the exponential piece is no longer arbitrary but fixed by the attractor
relation~(\ref{eq:Lambda_attractor}). Since the Hubble scale decreases as the
Universe expands, the post-inflationary evolution relevant for the mechanism
takes place in an increasingly well-controlled regime.

A separate important question concerns the anomalous coupling of the axion to
QCD, which is already present in~(\ref{eq:lagrangian_attractor}). This sector
is the same as in the non-compact QCD axion setup of~\cite{Karananas:2025uhy},
and its consistency is independent of the point discussed above, namely that
$F<H_I$. 

The tree-level unitarity of axion-mediated processes involving gluons
requires~\cite{Karananas:2025uhy}
\be
H_I \lesssim  \frac{32\pi^{5/2}}{g_s^2}\,f_a \ ,
\ee
or equivalently
\be
\label{eq:fa_bound}
f_a \gtrsim \frac{g_s^2}{32\pi^{5/2}}\,H_I \ .
\ee
Therefore, $f_a\sim10^9~{\rm GeV}$ is perfectly compatible with inflationary
spectator dynamics, but only for inflationary scales around
$H_I\sim10^{12}~{\rm GeV}$ or below; for $H_I\sim10^{13}~{\rm GeV}$, one
typically needs $f_a \sim \mc O ({\rm few})\times10^{9-10}~{\rm GeV}$. 

This has an immediate phenomenological consequence. The axion abundance is
controlled only by $F$, see~(\ref{eq:Omega_numeric}), whereas the residual
strong-CP violation scales as $\theta_{\rm phys}\propto F^{1/3}f_a$,
see~(\ref{eq:theta_phys_final}). On the other hand, consistency of the QCD
sector during inflation imposes the lower bound~(\ref{eq:fa_bound}) on $f_a$.
Therefore, increasing the inflationary scale forces $f_a$ upward. If one then
also requires the residual CP violation to remain below the experimental
bound, $F$ must be correspondingly decreased. Since the axion abundance grows
with $F$, this implies that increasing $H_I$ lowers the maximal dark matter
fraction that can be carried by the axion.

In this sense, high-scale inflation, perturbative control of the QCD sector
during inflation, and a small residual $\theta_{\rm phys}$ cannot be chosen
independently: together they constrain how large the axion contribution to the
observed dark matter abundance can be.

Let us note that reheating may give more stringent constraints, for both the
exponential, and the coupling of the axion to gluons; such an analysis lies
beyond the scope of this work.

\section{Origin of exponential potential}
\label{sec:exp_origins}

Before concluding, let us discuss a microscopic origin of the runaway
potential~(\ref{eq:exp_potential}), see also~\cite{Karananas:2026pov}. Because
the axion is taken to be classically shift-symmetric, it admits only a finite
number of couplings. In the matter sector, such an operator compatible with
the SU(3) gauge invariance is the one involving the axion and the QCD
topological density~$G\widetilde G$.

With gravity included, however, there are two further candidates on the same
footing, in that---just like the ordinary axion coupling to $G\widetilde
G$---they do not affect the dynamics perturbatively. Rather, their possible
importance is tied to nonperturbative gravitational effects. These terms are
\be
\mc L_{\rm grav} = \f{a}{F}E_4 + \f{a}{\widetilde F} P_4 \ ,
\ee
with
\be
\label{eq:gravity_densities}
E_4 = \f 1 4\epsilon_{\a\beta\gamma\delta} \epsilon^{\m\n\rho\s}
R^{\a\beta}_{~~~\m\n} R^{\gamma\delta}_{~~~\rho\s} \ , 
~~~P_4 = \f 1 2 \epsilon_{\a\beta\gamma\delta} 
R^{\a\beta}_{~~~\m\n} R^{\gamma\delta\m\n} \ , 
\ee
the Euler/Gauss-Bonnet and Pontryagin densities, respectively. In the above,
$F$ and $\widetilde F$ are arbitrary parameters of mass-dimension one, that
cannot be fixed by symmetry arguments. 

Despite their apparent structural similarity, the two operators
in~(\ref{eq:gravity_densities}) are expected to contribute in qualitatively
different ways once nonperturbative gravitational dynamics is taken into
account. This distinction is visible by analytic continuation to Euclidean
signature. The Euler density contains two Levi-Civita symbols and therefore
any contribution will be real. On the other hand, the Pontryagin density
contains a single Levi-Civita symbol and contributes an imaginary phase.
Hence, gravitational saddles carrying nonzero Euler and Pontryagin numbers
should lead to two structurally different contributions to the effective
potential for the axion; in the regime where a single saddle dominates, one
expects that 
\be
\label{eq:grav_potentials}
V_{E_4}(a) \sim \Lambda^4 \, e^{-a/F} \ ,
~~~
V_{P_4}(a) \sim \Lambda_P^4 
\cos\l(\f{a}{\widetilde F}+\theta_{\rm grav}\r) \ ,
\ee
where $\Lambda$ and $\Lambda_P$ control the strengths of the respective
potentials, and $\theta_{\rm grav}$ denotes a possible gravitational
$\theta$-angle.

Notice that the ``Euler sector'' naturally generates exactly the type of
exponential potential~(\ref{eq:exp_potential}). In this sense, the runaway
potential used in this work can be loosely interpreted as the low-energy
imprint of a shift-symmetric axionic coupling to gravity.

It should be stressed that for the mechanism to operate as described, the
``Pontryagin sector'' must be subdominant in two distinct senses. First, it
must be subdominant compared to the Euler one over the field range traversed
by the axion between the onset of the attractor and its eventual capture by
the QCD potential; otherwise the attractor dynamics would be modified. Second,
after QCD becomes important, the gravitational CP-odd contribution must remain
subdominant compared to the QCD potential itself; otherwise, the relaxation of
the physical strong-CP violation would be spoiled.

These are additional assumptions of such a gravitational origin for the tilt.
The same is true for the value of $F$, as its magnitude is determined by the
coupling and must be fixed to lie in the phenomenologically viable range
identified in the previous sections.

Let us finally point out that the possibility of gravitational origin for the
non-QCD contribution to the axion potential fits well to the Weyl-invariant
Einstein-Cartan construction of~\cite{Karananas:2024xja}, see
also~\cite{Karananas:2025ews}. In that setup, the axion is already part of the
gravitational sector, and a non-QCD contribution---a mass term---responsible
for lifting the degeneracy of the QCD branches is present at tree level. The
discussion above suggests a complementary possibility, \ie that
nonperturbative gravitational effects associated with the Euler density can
generate an exponential contribution of the form used here. If, over the field
range relevant for the post-inflationary evolution, the exponential
contribution dominates over the tree-level mass term, then the
attractor-driven cosmology described here is realized. Thus, the two scenarios
(statistical averaging~vs.~dynamical erasure of isocurvature perturbations)
can be viewed as different regimes of the same gravitationally-motivated
framework.

\section{Conclusion}
\label{sec:conclusion}

We have discussed a dynamical mechanism for erasing inflationary isocurvature
perturbations of the QCD axion. The key ingredient is a runaway exponential
potential, which drives the axion onto the scaling attractor well before the
QCD epoch. Once on the attractor, the axion tracks the dominant background and
forgets its inflationary initial conditions, so that the would-be isocurvature
mode is strongly suppressed. 

When the QCD potential turns on, the fate of the field depends on how much
kinetic energy it has acquired on the attractor relative to the height of the
barriers of the periodic potential. If the exponential is steep, the field
arrives moving fast enough to ride over several barriers before finally being
trapped. This is the kinetic-misalignment regime, in which the abundance is
governed by how long the field keeps rolling. Obtaining all of the observed
dark matter points to a slope characterized by $F\sim \mc O (10^{11})~{\rm
GeV}$. If, on the other hand, the exponential is ``gentle,'' the field arrives
slowly and gets captured immediately. It then behaves as in ordinary
misalignment, and its abundance is set by the QCD sector alone, \ie the decay
constant. 

The same exponential tilt that makes the attractor possible also shifts the
minimum of the axionic potential away from the CP-invariant point, inducing
residual strong-CP violation in both regimes. In the kinetic-misalignment
regime, where the dark-matter abundance fixes $F$, this ties together the
inflationary scale, the lower bound on $f_a$ from consistency of the QCD
sector, the maximal axion dark-matter fraction, and the size of $\theta_{\rm
phys}$, yielding a residual phase close to current bounds and a target for
electric dipole moment searches.

A key phenomenological distinction from our earlier statistical non-compact
mechanism~\cite{Karananas:2025uhy,Karananas:2026pov} is that no domain wall
network is formed here. The attractor homogenizes the field well before the
QCD potential becomes relevant, and the observable Universe is captured into a
single branch. The only metastability issue is then quantum tunneling to the
neighbouring minimum, and we have shown that the corresponding decay rate is
extremely small. We stress, however, that the mechanism selects a branch
dynamically without addressing why that branch has the observed vacuum energy,
so the cosmological constant problem is not addressed.

\section*{Acknowledgements}

We are grateful to Sebastian Zell for comments on the manuscript.

\bibliographystyle{utphys}
\bibliography{Refs.bib}

\end{document}